# Elucidation of the origins of HTSC transport behaviour and quantum oscillations

**John A. Wilson**

H. H. Wills Physics Laboratory
University of Bristol
Tyndall Avenue
Bristol BS8 1TL.  U.K.

**Abstract**

A detailed exposition is made of recent transport and 'quantum oscillation' results from HTSC systems covering the full carrier range from overdoped to underdoped material.  This now very extensive and high quality data set is here interpreted within the framework developed by the author of local pairs and boson-fermion resonance, arising in the context of negative-$U$ behaviour within an inhomogeneous electronic environment.  The strong inhomogeneity comes with the mixed-valent condition of these materials, which when underdoped lie in close proximity to the Mott-Anderson transition.  The observed intense scattering is presented as resulting from pair formation and electron-boson collisions in the resonant crossover circumstance.  The high level of scattering carries the systems to incoherence in the pseudogapped state, $p < p_c$ (= 0.183).  In a high magnetic field the striped partition of the inhomogeneous charge distribution becomes much strengthened and regularized.  Magnetization and resistance oscillations, of period dictated by the favoured positioning of the fluxon array within the *real space* environment of the diagonal 2D charge striping array are demonstrated to be responsible for the recently reported behaviour hitherto widely attributed to the quantum oscillation response of a much more standard Fermi liquid condition.  A detailed analysis embracing all the experimental data serves to reveal that in the given conditions of very high field, low temperature, 2D-striped, underdoped, $d$-wave superconducting, HTSC material the flux quantum becomes doubled to $h/e$.

**§1. Background to negative-*U* modelling of HTSC phenomena.**

It long has been realized that the normal state properties of HTSC cuprates, whether electrical, magnetic or optical, are highly unusual, and bear an intimation of what is to follow within the low temperature superconducting condition [1]. All the normal state properties give indication of chronic electronic scattering, especially along the axial Cu-O planar bond direction; *i.e.* in the antinodal direction of the superconductive *d*-wave state and containing the saddle points of the Fermi surface and their so-called 'hot spots' (the latter associated with the oblique diagonals of the B.Z. quadrants; see fig. 3 in [2]). This unusual scattering becomes increasingly manifest as the employed 'doping' level is lowered through the overdoped condition towards $p^{opt}$, and to it is attributed the steady rise with cooling in the magnitude of the observed *p*-type Seebeck [3,4] and Hall [5] coefficients toward non-metallic levels.

ARPES was the first technique to reveal directly that this scattering-induced degradation in Fermi behaviour sets in from the zone-edge saddles and slowly develops round the Fermi surface (FS) in the direction of the 45° zone diagonals as both *p* and *T* fall [6,7]. This impairment of band-like quality (i.e. induced incoherence) that the carriers undergo with decreasing hole 'doping' does not initially see a diminution in $T_c(p)$ but, rather, an increase. However, as long known from electronic specific heat studies [8], somewhat in advance of reaching $p(T_c^{max})$ the superconducting condensation energy suddenly collapses. From my long-standing perception of HTSC phenomena as being the outcome of a resonant boson-fermion crossover in an inhomogeneous (mixed-valent) negative-*U* setting [1,9], I regard this collapse to be the moment FS disintegration finds coherent quasiparticle status extinguished at the hot spots. (The latter are the *k*-points from which the doubly-loaded negative-*U* states ($^{10}Cu_{III}^{2-}$) are most favourably accessed [2]). There then ensues a rapid development and extension in the density-of-states 'pseudogap' to lower *p*.

In real space I have indicated in fig. 5 of [10], within the dynamic stripe phase context of LSCO, that the moment at which maximum condensation energy arrives comes in close association with a 7x7$a_o$ 'stripe' structure and doping content *p* = 0.183. At this stage as the key carriers are driven close to their band-like limit they are rendered best able to form and support long-lived local pairs within the $Cu_{III}$ clusters of site-charging which the 2D striping geometry presents. A 7x7 structure is, note, spatially compatible with the given hotspot location. Mention of $Cu_{III}$ clusters emphasizes here that the negative-*U* state is associated not with the copper cations *per se*, but with entire Cu-O coordination units. The key to the negative-*U* effect is the $p^6d^{10}$ shell closure achieved with the $Cu_{III}$ site charge double-loading $^{10}Cu_{III}^{2-}$ and ensuing total reordering of the till now strongly $\sigma\sigma*$ bonding and antibonding O(*p*)/Cu(*d*) states [9].

The local pair states being so generated are most effective at elevating $T_c$ when they are neither too localized (or few in content) as at low *p*, nor too short-lived and weakly interactive as under the well screened conditions of high doping. In the desired crossover conditions of the fermion-boson mix, the local pair bosons need to uphold within the Fermi sea as many ancillary BCS-like quasiparticle couplings as possible. The optimal of optimum conditions at raising $T_c^{max}$ is acquired by selecting that material for which the negative-*U* state binding energy ($U$) below $E_F$ stands resonant with the induced chemical potential at the very moment set by the geometrical condition above and completion of the DOS pseudogapping at the hot spot. Control in achieving these optimized circumstances is granted within the present scheme by a manipulation of the ionicity of the counter-ions incorporated into the system. $HgBa_2Ca_2Cu_3O_{16+\delta}$ currently constitutes the most favourable combination. What the selected level of covalency of the overall system permits is close control over the metallic screening operative within the material and of the dielectric constant to follow. If a system is too covalent (as say $Bi_2Sr_2CuO_{6+\delta}$), metallic screening remains too active at $p = 0.18$ and the population/lifetime of the local pairs is diminished. If the system is too ionic (as say $(Na_xCa_{2-x})CuO_2Cl_2$) then by $p = 0.18$ the quasiparticle incoherence will have become too advanced and the number of induced BCS-like pairs that can be sustained is lessened. Note a small degree of gain in stability for the local pairs can be tolerated at the expense of some minor loss in band carrier number. Indeed $T_c^{opt}$ uniformly is to be found at $p = 0.16$, not $p = 0.18$, but equally uniformly remember that the superconductive condensation energy per Cu atom maximizes at the latter concentration within all cuprate HTSC systems [8].

    The systematics of the connection between the final ($d$-wave) superconductive gapping parameter $\Delta_o^{max}$ and $U$, the negative-$U$ state binding energy, when arrayed as functions of doping and system covalency has been sketched out in figure 4 of [11] using information drawn principally from energy- and position-resolved scanning tunnelling microscopy, from ARPES and from neutron scattering. The scheme presented in [11] has very recently received added support from the $p$-dependent STM results for Bi-2212 just published by Kohsaka, Davis and colleagues [12]. Needless here to say, I continue to depart strongly from the interpretation which the latter team offer of their results, although it is good to see now the nodal/antinodal dichotomy being acknowledged in relation to the two principal energies involved in the phenomena. (Beware when comparing fig. 3b in [12] with my figure 4 in [11] that in the former the modes plotted for the four different $p$ values have received increasing zero offsets). Kohsaka *et al* fail to perceive why the features they record exist only over a limited range of angles $\phi$ around the Fermi circle. Within my understanding this arises because the mode in question, to me the dispersed uncondensed pair mode [9c], is defined sharply only between the hot spots and where this linearly dispersed mode emerges above $E_F$. Note that while the induced nodal superconductivity is demonstrably of $d_{x2-y2}$ $B_{1g}$ symmetry, the local pair state takes angularly empathetic but distinct, extended-$s$, $A_{1g}$ form. The local pairs become unstable near the zone centre, and in truth in the STM data there is no indication that the observed modal feature pulls round in simple $d_{x2-y2}$ fashion toward the gap node. Note all recorded very low energy investigations in HTSC materials, such as electronic specific heat, penetration depth, $B_{2g}$ electronic Raman and optical measurement, are those concerning the induced BCS-like pairing.

    §2 now tracks in detail the development of the abnormal electrical and optical properties of HTSC systems as 'hole doping' is reduced below $p = 0.3$. For the latter the position of the Landau Fermi liquid behaviour is fairly standard if highly correlated. Below $p = 0.3$ there arises marked growth in scattering and a steady shift towards incoherence. Apportionment of the

unusually strong scattering into elastic and inelastic, isotropic and anisotropic contributions is presented in terms of *e*/*e*-to-*b* and *e*-on-*b* events from a close examination of the *T*- and *p*-dependent evolution of the AMRO, Seebeck and Hall data.

§3 takes a detailed look at the new high quality $\rho(T,p)$ resistivity data and analysis from LSCO, and at the changes that suddenly occur at $p = p_c$ (=0.18). The nature of what arises there as one encounters strong incoherence and the pseudo-gap condition is fully described in terms of the crossover model.

§4 follows up on the consequences of such an interpretation and attempts to reformulate the setting of the high-field oscillatory magnetization, resistivity and Hall data recently obtained from UD material in terms *not* of the standard quantum oscillation, *k*-space response of a fairly normal Fermi liquid, but of a real-space-based action much more appropriate to the UD condition. Success here is demonstrated to revolve around the 2D striping supported by such mixed-valent materials, and seemingly enhanced in definition by the strong applied magnetic field. The model is one of response to the free energy changes arising as the unpinned vortex array slips discretely past the doping determined stripe array. A very close match to the experimental data can be secured *provided* that the flux quantum in question in the ring-threading process is not *h*/2*e* but *h*/*e*. The observations are related back to Forgan and colleagues' SANS observations on UD material of a transfer from hexagonal to square symmetry in the vortex array geometry as the magnetic field is increased.

## §2. The nature of the scattering governing HTSC transport and optical behaviour.

Following this introduction I wish now to examine in detail the important recent work from my colleagues in Bristol relating to such matters, work in the main performed under strong magnetic fields in excess of 30 tesla. Their dc and ac resistivity data [13,14], their Hall data [15], and their now renowned Fermi surface quantum oscillation data [16,17,18] all afford great potential insight into the above matters. Unfortunately it must be said that it has not proved possible to get them as yet to commit themselves with regard to the HTSC mechanism. Since I have long since jumped, it is incumbent upon me here to offer interpretation along the above lines of the phenomena which they are reporting. In this it is best to proceed following roughly the order in which the data were acquired, starting with the AMRO (angular magneto-resistance oscillation) results on overdoped $Tl_2Ba_2CuO_{6+\delta}$ (Tl-2201) [16].

Tl-2201 is a very valuable system within the array of HTSC materials in that it is structurally simple (bar some cation cross-substitution), is of relatively high $T_c^{max}$ (~ 95 K), and, above all, may be taken from optimal hole doping right through into the range beyond $p = 0.28$ where superconductivity has vanished. On moving to somewhat smaller *p* than this, one can stay within the normal state and study the low temperature condition there by the application of magnetic fields in excess of $H_{c2}(p,T)$. Under these conditions the AMRO work delivered the first low energy, low temperature record of a relatively well-formed Fermi surface in highly overdoped material. Of course, as the level of *p* doping is reduced, $T_c$ quickly picks up and $H_{c2}(p)$ passes above 50 tesla, thereby limiting further investigation. However it was very apparent even from the limited range accessed that scattering in the system is mounting rapidly as $p$ is reduced, bringing down the $\omega\tau_c$ values extracted from the data to below unity. The most significant observation made in the analysis of this AMRO work was that the abnormal intense scattering is comprised of an elastic, anisotropic ($\propto v_F^{-1}$), temperature-independent base term and two temperature-dependent terms, one characterized by $T$-linear behaviour and strongly anisotropic in form (this, like the base term, is maximal in the axial saddles) and the other isotropic but inelastic and characterized by a $T^2$ variation.

$$\Gamma(T,\phi) \;=\; \{\; \Gamma_0(\phi) \;+\; \Gamma_1.\cos^2(2\phi).T \;\} \;+\; \Gamma_2.T^2$$

|  | anisotropic | isotropic |
|---|---|---|
|  | elastic    inelastic | inelastic |
| – ascription made in present text: | electron-boson | electron-electron to boson . |

These temperature behaviours extend up to such values as indicate their sourcing to be purely electronic and not phononic. Normally $T^2$ $e$-$e$ Baber scattering is in evidence only at very low temperatures, prior to being swamped by $e$-$ph$ scattering, but clearly the $e$-$e$ scattering now in play must be super-effective for the term to remain in evidence at all temperatures. In [1,19] I submitted that such a $T^2$ term must represent local-pair boson formation. It comes to dominate nodal scattering. How long the pairs persist will depend upon the $p$-value and the level of screening supported (although not overly upon the temperature once above $T_c$, as apparent from laser pump-probe work [19]). Naturally when screening falls as $p$ is reduced, these bosons acquire a longer lifetime and the population of pairs, both local and induced, mounts steadily as $p$ drops towards $p^{opt}$ at 0.18. By that stage one can anticipate $e$-$b$ scattering is going to predominate, the $T^2$-term to the strong scattering now yielding primacy to the $T$-term.

The $e$-$e \rightarrow b$ isotropic scattering in its strongly inelastic character reflects its third party nature, the lattice needing locally to accommodate (swell) under the $Cu_{III}$ site, charge double-loading event. Inevitably here within a highly correlated system the electrons cannot be treated in isolation from the lattice, as made very apparent from the isotope effect data [20], the neutron-acquired soft mode data [21], the reported transient laser-induced changes in lattice parameter [22], and indeed from direct lattice and bond length measurements themselves [23,24]. In contrast to the above, once the population of axially stabilized bosons is extensively acquired, under $p$ levels to support $T_c^{max}$ and $E_{cond}^{max}$, the then dominant $e$-$b$ scattering is anisotropic and quasi-elastic. Each scattering boson naturally is unrestricted here by the Fermi principle with regard to the $k$-states into which it is able to pass. Hence whilst this electron-on-boson scattering is very intense because of the augmented scattering cross-section, it predominantly will be small-angle in form. As was previously expressed in [9b], I do not see the source suggested by Varma and Abrahams [25] for this small angle scattering in HTSC materials as being realistic; namely a scattering off the 'defects', intrinsic and otherwise, residing between the $CuO_2$ layers. Why then (i) the observed very strong thermal augmentation; (ii) its occurrence in OP Y-123 and in lattice perfect Y-124; (iii) its super-strong showing exclusively for HTSC materials, and a relative absence for partially intercalated layer compounds such as $Na_xTiO_2$ or $Cu_xTiS_2$? The current scattering, as expounded upon recently by Zaanen [26], is taking the HTSC materials to the verge of incoherence, with $h/\tau \rightarrow 2\pi kT$. It is occurring, what is more, in materials which, as a result of the advanced *p-d* mixing in hole-doped cuprates, display a mean-field LDA bandwidth for the FS-containing $d_{x^2-y^2}$ sub-band manifesting the not insubstantial value of 2 eV plus [27].

Everything recorded about HTSC materials implies Marginal Fermi Liquid (MFL) behaviour of some form [25]. Hence one is obliged not to allow the recent observation of dHvA oscillations in very strongly overdoped Tl-2201 [13] to obscure this basic aspect once one enters the HTSC range of *p*. In particular this becomes the case for underdoped Y-123 and Y-124 in regard to the purported revelation of small, rather ordinary FS pockets from SdH and similar data [18] – of this much more later. HTSC materials are far from being ordinary metals as any glance at the low temperature infra-red results will reveal (with their sizeable mid-IR oscillator term, residual temperature-dependent Drude term, and scattering $h/\tau \sim 2\pi k_B T$) [28]. Even ignoring matters specifically involving the l.t. condition, the proximity of the cuprates to Mott insulation (extant in all *single*-valent $Cu_{II}$ and $Cu_{III}$ oxides) plus the strong local disorder, electronic as structural implicit with the mixed-valence of the HTSC materials, ought immediately to direct one away from simplistic considerations of what is afoot. Nowhere is this more in evidence than when confronting the Hall data for HTSC systems. Although the (300 K) Hall constant sign is positive, as might well emerge from a Fermi surface centred upon the zone corner, its magnitude is remarkably high at low substitution levels away from band half-filling, a fact that has greatly contributed with the HTSC cuprates to the whole notion and terminology of hole 'doping' *p* away from the $d_{x^2-y^2}$-based Mott state, $^9Cu_{II}^0$. The task set is to tread warily through the complex region between band and localized behaviour without any lapses of concentration, such as those I would claim have arisen in much discussion of the recent SdH/dHvA results from underdoped material ($p \sim {}^1/_8 - {}^1/_{10}$). The low energy Hall [5,15] and Seebeck [3,4] data afford the most favourable point of entry to penetrating this crossover behaviour.

One of the early-noted empirical rules established by Obertelli, Cooper and Tallon [4] was that in every HTSC system the Seebeck coefficient for optimally doped material always on warming changes sign from positive to negative at just about 300 K. The early *ad hoc* analyses of this temperature dependent Seebeck coefficient invariably revolved around standard gaps. The latter were universally determined as being extremely small (« 50 meV), and in [3b] I indicated that this was not an appropriate model with which to proceed. Informed by the early electronic specific heat analysis from Loram and coworkers [8] and the formal treatment of the Seebeck coefficient provided by Hildebrand *et al* [29], we in [3a] presented a treatment of such data from Hg-1201 based upon a negative-*U* approach with density of states pseudogapping in the presence of resonant B-F scattering. In the thermoelectric expressions the observed positive sign of *S* is engendered by the sign at $E_F$ of the gradient $\partial\sigma/\partial E$ being negative, which in turn issues from $\partial\tau/\partial E$ and/or $\partial n_{eff}/\partial E$ being negative. In optimally doped systems such a state of affairs terminates by 300 K, but with underdoped systems the pseudogapping onset temperature continues to mount, a feature now well-monitored in STM and ARPES work, and the Seebeck coefficient stays positive to raised *T*. Note this many-body pseudogap condition becomes lost towards *T** not via reduction in gap energy but by DOS 'in-fill' – hence its name.

With the Hall coefficient the analysis of the conversion from positive to negative sign is not quite so straightforward, and crossover does not occur at the same $\{p,T\}$ combination as for the Seebeck coefficient. This is because the Hall process is a transverse one, it, as demonstrated by Ong [30], being dependent upon the cross-product, Stokes area integral in $k$-space at $E_F$ – $\int_{FS} \mathbf{dl} \times \mathbf{l}$ relating to the local mean free path vector $\mathbf{l}$ at each particular $\mathbf{k}_F$. The different weighting in the integral coming from different bits of Fermi surface is, in addition to anisotropy in $\mathbf{v}_F$, then dependent upon the anisotropy of the scattering rate $1/\tau(\mathbf{k}_F)$ (*i.e.* on $\mathbf{l}(\mathbf{k}_F)$), and this critically so when that scattering is super-strong and highly anisotropic, as in the present case. The outcome is that, for many HTSC materials, $R_H$ remains positive to very high temperatures and also to surprisingly high $p$ values – indeed even to beyond where in LSCO the band saddle-point has been surmounted and the FS become closed about the zone centre rather than the zone corner.

Figure 1 illustrates the latter circumstance for OD LSCO, drawn appropriate to the level of anisotropy $\mathbf{l}(\mathbf{k}_F)$ disclosed by the AMRO work on OD TBCO, following analytical treatment which, as long as coherence is fully maintained, remains able to assume standard Boltzmann-Zener-Jones type form. At $p = 0.3$ we are looking at strong anisotropy of about 3.5 in basal $\mathbf{v}_F$, while for $|\mathbf{l}|$ (or scattering rate $1/\tau$) a weak anisotropy at 300 K of about 10% remains. As $p$ and particularly $T$ are reduced the evaluated anisotropic (*e*-on-*b*) contribution to $1/\tau$ steadily mounts. Ultimately there has to come an end to the validity of the B-Z-J analysis somewhere in advance of $p = 0.183$, the value signalled above as marking the quasiparticle coherence limit at the hot spots. Nevertheless the modelling presented in figs.3 and 4 of [15] by Narduzzo *et al* of the rising magnitude of $R_H$ under falling $p$ and $T$ was quite promising. At the time in fact only the $T^2$ term and the band anisotropy in $\mathbf{v}_F$ were inserted into the analysis, suspecting Fermi liquidology to prevail. The outcome clearly was, though, to understate the rise in $R_H$ as $p$ fell. What could not, moreover, be captured was the growth upon cooling as the local pairs move into operation of actual precursor superconductivity within micro-regions where $\sigma \rightarrow \infty$ (*i.e.* $V_{RH} \rightarrow 0$). From low-field work one observes well in advance of $T_c$ itself that the Hall coefficient, as with the Seebeck coefficient, begins to disclose its necessary decline towards zero at $T_c$, this turn around being encountered the earlier the more underdoped the sample is, *i.e.* the greater $U$ and $T^*$. Such behaviour parallels what is in evidence in the Nernst data above $T_c$ [31,9c].

To secure a truly satisfactory outcome to all these scattering calculations, whether when handling under- or overdoped HTSC data, Hussey *et al* [32] realized some time ago that it is necessary to supplement application of the scattering equation above somewhat further. In consequence of the strongly temperature dependent terms there, the average scattering mean free path quickly is reduced to approach the Mott-Ioffe-Regel limit of $l \approx a$ (the lattice parameter). This marks the coherence limit to band-like quasiparticle behaviour ($\mu \sim 1$ cm$^2$/V-sec). It affords a ceiling within many heavy scattering materials, such as the A15's, against an unending escalation in the overall scattering rate, in what has become known as 'resistance saturation' [33]. When scattering is highly anisotropic the approach to saturation inevitably will reflect the strongly directional nature of $|\mathbf{l}(\mathbf{k}_F)|$ around the Fermi surface. The situation calls for an added resistance-limiting term to obtain what rather unfortunately has become known as the 'parallel resistance' formulation. In this the effective net scattering rate $\Gamma_{effective}$ ($\propto 1/\ell_{eff}$) is secured from the original expression, $\Gamma_{ideal}$, via the inclusion of a saturation scattering term $\Gamma_{maximum}$ such that $1/\rho_{eff} = 1/\rho_{ideal} + 1/\rho_{max}$ : I use here the terminology employed by Hussey in [32]. The expression is effectively one of conductances acting in sequence and relates to coherent and incoherent micro-segments within the overall transport process. In order to control the number of free parameters in the numerical work, the angular and $p$ dependence of $\rho_{max}$ invariably has been suppressed, a universal value being inserted for $\rho_{max} \sim 1$ m$\Omega$-cm ( $\equiv$ $h\Gamma_{max} \sim 3,000$ cm$^{-1}$ or 0.4 eV). Despite complete saturation in OD-Tl-2201 not being encountered until 600 K, this extra term is witnessed to make its presence felt at much lower temperatures.

As states become incoherent (*i.e.* lose their band-like quality) they will relinquish any related *p*-type signing in regard to the contribution they make to the Hall coefficient (although note amorphous and glassy metals can be *p*-type [34]). Prior to actually being lost to the band-like cohort the near-axially directed quasiparticles are of very low mobility and despite being great in number the contribution they make to $R_H$ is not here dominant. Recall in Hall work the quantity the Hall mobility, $\mu_H = R_H.H_z/\rho$, is often introduced. In the present case $\mu_H$ will be dominated by those carriers moving in the less severely scattered nodal directions. Such carriers are primarily suffering the (isotropic) $e$-$e \rightarrow b$ based scattering, $T^2$ in form. Accordingly $\mu_H$, or its equivalent $\tan\theta_H/H_z$, will overall be proportional to $T^{-2}$ [1]. Indeed the empirical relation $\cot\theta_H = A + BT^2$ long has formed one of the best recognized elements of HTSC phenomenology [35]. We see now this ubiquitous (approximate) form occurs by virtue of the very strongly anisotropic nature to the electron-boson scattering, high in the saddles and low in the nodal directions. Within the diagonal stripe modelling of [10] note, in addition, that the nodal direction supplies the easy 'rivers of charge' direction. It might well be that the $T^2$ scattering term indeed actually develops some anisotropy of its own as $p$ and $T$ fall, but holding this contrasting B$_{2g}$ geometry to the B$_{1g}$ form of the $T$ term.

For those who haven't yet adjusted to 'nodal rivers of charge', I advise they look again at the outcome of Homes *et al's* sub-60 K infrared study on $p = {}^{1}/_{8}$ LBCO (wherein $T_c$ is much depressed) [28b]. At such $T$ and $p$ many carriers, as apparent from the dip in the IR reflectance edge, have become rendered incoherent and ineffective at contributing towards the Drude optical response. As *axially* oriented carriers become incoherent, those quasiparticles left in the nodal direction actually become less strongly scattered and the residual Drude peak indeed sharpens up at the lowest temperatures. Such carriers running freely in the nodal directions, and maintaining the Fermi arc signal of ARPES work [6], convey a 2D behaviour completely at odds with the much vaunted 1D setting to stripe formation and activity (see [10]). Not only does the $T^2$ scattering behaviour characterize the nodal response, but it is echoed too optically in a nodal self-energy that varies as $\omega^2$ [36]. Aeppli *et al* [37] many years ago in fact noted that the high energy, inelastic neutron scattering $\pi,\pi$ 'resonance' linewidth manifests a joint dependency upon $\omega$ and $T$ proceeding as $\surd(\omega^2 + T^2)$. I close this section by drawing attention to the fact that in *LBCO* under the above conditions the Hall coefficient is not in fact positive but negative [18c,d], a matter of some significance to which we shall return.

Let us before that turn again to the Bristol-based transport work of Hussey and colleagues. While Hussey proceeds in his review of [32b] to deal next with the very unusual magneto-resistance data over the matter of their non-Kohler-like behaviour within the framework already developed, we are in a position now here to address directly their most recent high-field resistance results from high quality, single crystal samples of optimally and overdoped LSCO [13]. This new data stands sufficiently accurate and self-consistent between the different samples to permit a really close investigation into how the scattering progresses with reduction in *T* and *p*. The new paper in fact has been couched within the presently popular setting of quantum critical phenomena, but the outcome is far from what some had anticipated, and it provides a most revealing view of what actually is underway within HTSC materials.

**§3. Detailed look at new resistivity and Hall data in relation to HTSC mechanism.**

The new data [13] come from a sequence of eight different, well-specified compositions of LSCO single crystal with *p* running from 0.17 to 0.33, and were acquired under z-axis magnetic fields of up to 60 T. The quality of the data is such as to make it possible meaningfully to fit to the full scattering procedure above and to extract the *p*-dependent behaviour for coefficients $\alpha_1$ and $\alpha_2$ relating respectively to the *T* and $T^2$ terms. This requires being able satisfactorily to extrapolate the observed $\rho(H,T,p)$ plots down into the superconducting regime below the relevant $H_c(T,p')$ values so as to uncover the l.t. condition that would prevail in the absence of LRO superconductivity. The extrapolation has too to traverse appropriately the SRO fluctuation range above $H_c\backslash T_c$. What is found is that right up to *p* = 0.33 *both T* and $T^2$ terms are always present, but that their relative weighting changes most illuminatingly both with *p* and *T*.

As indicated above some among the authors of the work had expected to be able to link their findings into the general discussion of quantum critical phenomena, after the fashion of various rare-earth systems. There a linear-in-*T* behaviour for $\rho$ is associated with a rather narrowly divergent fan on the *T vs. x* (or *P*) phase diagram, flanked by $T^2$ Fermi-liquid behaviour. However, one observes for the HTSC results that both terms in fact run concurrently throughout the entire superconducting range of *p*, and, moreover, that the (anisotropic) linear-in-*T* term, instead of fanning out to high *T*, is confined in the main to low temperatures for which pairs are to be found, *i.e.* onsetting around *T\** and dominant below $T_c$. As claimed before this particular term is surely the expression of *e*-on-*b* scattering. If one proceeds with a single-term empiric formulation $\rho \propto T^n$, as has often been done in the past, then exponent *n* drifts from 2 towards 1 as *T* and *p* fall and the life-time and instantaneous population of bosons build.

Interestingly the analysis, as conducted, does not manage to pick up any augmentation in *e-e* → *b* pair production success with fall either in *T* or *p*, as gauged by the discerned behaviour of $\alpha_2$ for the $T^2$ term. There occurs no detectable change in this coefficient right from *p* = 0.33 down to $p_c$. By contrast the evaluated $\alpha_1$ (*e*-on-*b*) coefficient picks up steadily with reduction in *T* or *p*, as the population of bosons rises. Once again the latter behaviour continues through until $p$ drops to $p_c$. So what is this $p_c$? It is 0.183, the value of doping for which the superconducting condensation energy per carrier is at a maximum [10], as revealed by the specific heat analysis of Loram and coworkers [8]. The combination of boson population and boson binding energy has become optimized at this juncture. As stated already, it is with this concentration one seriously is encountering incoherence in the quasiparticle system. Below this hole concentration quasi-particles rapidly become abstracted from the coherent response of the system and the condensation energy falls away sharply. While the Uemura plot, made from muon penetration depth work [38], had revealed from low $p$ the steady build up in pair numbers $n_s$, the effectiveness of those pairs in advancing the global condensation energy was, it is evident, somewhat less than their effectiveness at raising $T_c$. From $p_c$ down into this sub-optimally doped region, where the local pairs are dropping out of resonance with $E_F$, the energetics of the coupling between the two subsystems, their fermions and bosons, is declining [11], and at low $p$ $\rho$ returns to dominantly quadratic form [39].

So how is this manifest in the new resistivity results as one proceeds below $p_c$? $\alpha_1$, having come to a broad peak at $p_c$, now falls away quite steeply. Conversely $\alpha_2$, so long $p$-independent, exhibits a sharp upward movement below $p_c$, bringing a higher absolute and relative contribution to $\rho$. By $T_c$ (≈ 33 K) the magnitude of $\alpha_1$ at $p_c$ is raising the value of $\rho$ to its coherence limit of 1mΩ-cm. As expressed previously, the downturn in $\alpha_1$ below $p_c$ declares a drop-off in the overall boson population if the linear-in-$T$ term indeed is registering $e$-$b$ scattering activity. $p_c$ stands at the tip of the Uemura 'butterfly wing', i.e. of the $T_c(p)$ vs. $n_s(p)$ plot [38]. Conversely with $e$-$e \rightarrow b$ scattering indeed gauged by the $T^2$ term, $\alpha_2$'s sudden rise below $p_c$ (for all $T$) must express the enhanced cross-section for local pair production once metallic screening starts to collapse upon the onset of large-scale quasi-particle incoherence. As noted by Hussey and coworkers [13] this counter-movement between $\alpha_1$ and $\alpha_2$ encountered at $p_c$ is the precise opposite of what might be expected were one looking at standard quantum critical behaviour. This truly is new physics – but not that physics.

The development of incoherent quasi-particle behaviour below $p_c$ immediately is picked up too in the room temperature Hall coefficient, which departing from being determined by the entire Fermi surface complement of quasiparticles above $p_c$ (as in the Narduzzo work [15]), now regresses to track only the number of 'holes' away from $^9Cu_{II}^0$ [5]. In the extreme stripe model the latter reside solely on the stripes, the rivers of charge [10]. The pseudogapping records this rapid decline in the number of active quasiparticles. Within the domains between the stripes the Cu sites there all move steadily towards a frozen $^9Cu_{II}^0$ Mott aspect. Such changes have a marked effect upon the $p$ dependence of the chemical potential, it transferring from band-like accommodation to $p$ (i.e. to the average Cu valence) above $p_c$ to becoming pinned in energy below $p_c$, as for a doped semiconductor. This crossover in behaviour can readily be followed by core-level photoemission [40]. In LSCO where striping is most extreme the transformation is very clear cut (see Hashimoto, fig 3). Note the states for which the strongest changes in core line position are manifest are the Sr and La 3d states, the ones most directly interactive with and responsive to any modification to the Cu 3d conduction band character. The loss of spectral weight from the band-like cohort of states, which loss of coherence and the pseudogapped condition entails, has recently also been pursued through analysis of the electronic Raman spectra [41] and the ARPES spectra [42] by Storey *et al* and Sahrakorpi *et al* respectively. One does not have to believe in Fermi arcs *per se*, but simply that persistence of the chronic scattering impels a broadening of their *complementary* states into incoherence; specifically the states of the F.S. saddles most active in the negative-*U* boson/fermion crossover events.

An additional observation relating to Hall work might be dealt with here before turning finally to the matter of quantum oscillations in underdoped HTSC material. This concerns the recent high-field (50 to 65 T) data from Balakirev *et al* [43] and the uncovering just below $p = p_c$ of *non-monotonic* behaviour in $R_H(p)$ at low $T$. The new work uses high quality thin-film samples of LSCO and is a follow-up study to comparable work on $Bi_2(Sr_{2-x}La_x)CuO_{6+\delta}$ in 2003. When the authors make to convert their $R_H(T,p)$ findings into an effective Hall number $n_H(T,p)$ via a simple inversion it might seem that the non-monotonic behaviour at low temperature amounts to a temporary recovery in active quasiparticle numbers after the initial fall at $p_c$. This recovery would peak narrowly a little above $p^{opt}$, before ultimately $n_H$ falls away steeply again. In [43] there follows once more an appeal to QCP for explanation of this perceived behaviour. However let us recall that the established anisotropy in $l(\mathbf{k}_F)$ renders simple inversion of $R_H$ for $n_H$ invalid. It is my understanding that what is being observed here emerges from the rapid improvement in the *average* mean free path following the elimination from play of the antinodal quasiparticles. It was revealed some years ago now by Krishana *et al* [44] from thermal conductance work that in a magnetic field the electronic contribution to $\kappa_{th}$ is increasing rapidly in general vicinity of $T_c$. The present Hall experiment for which $T_c$ is quenched would indicate this rise to be reliant not on precursor superconductivity but on the changes to the 'normal' state electronic *scattering* coming with such $p$ slightly below $p_c$.

### §4. An end to Fermi liquid quantum oscillations in underdoped HTSC material.

There is no problem with seeing quantum oscillations as set by the large Fermi surface present in strongly overdoped HTSC material, provided one is able to go to sufficiently high magnetic fields (> $H_{irrev}$). dHvA and SdH studies on strong coupling superconductors have become routine since the ground breaking work on 2H-NbSe$_2$ [45], even though $H$ might stand below $H_{c2}$. Very recently Vignolle *et al* [17] have managed to secure both magneto-resistance and magnetic torque oscillations from small, high quality (RRR~20), highly overdoped Tl-2201 crystals, with $T_c \approx 10$ K, by working at sub-4.2 K temperatures and in fields of between 50 and 60 tesla (here > $H_{c2}$). The single observed Fourier frequency of 18,100 tesla corresponds to a hole F.S. occupying 65% of the area of the Brillouin zone, *i.e.* to a '*p* value' (reckoned from half filling) of 0.30. That outcome is totally in accord with band structure calculation [46], with ARPES data [47], with the angle-dependent magneto-resistance data [16], and of course with the $T_c$ value [48]. No problem here then, but there occurs one alerting feature emerging from the data, namely the quasi-particle effective mass. Here at 4.1 $m_e$, $m^*$ is already very considerably greater than the calculated band mass of of 1.2 $m_e$, whilst the mean free path even in this highly overdoped material is only ~500 Å – and yet we still have to encounter the vast bulk of the correlated behaviour to come as the value of *p* is reduced.

It has been my longstanding belief that one cannot contemplate then results in any way comparable to the above coming from optimally and especially underdoped material. People for many years have endeavoured to obtain dHvA/SdH oscillations from optimally doped Y-123, Bi-2212, etc., including operating to much higher pulsed fields, but without any sign of success. So how could it be that several groups have over the past 2½ years detected what appear to be very like quantum oscillations issuing from - all the more remarkably - material that is significantly underdoped, namely $YBa_2Cu_3O_{6.5}$ [18c] and $YBa_2Cu_4O_8$ [18b], wherein, as we saw earlier, the chronic scattering has brought the Fermi liquid into incoherence? Clearly the answer to this riddle has to lie with the fact that these experiments on the underdoped material turn up a single frequency some 30 times smaller than what is seen with OD-Tl-2201. This would amount to an area in *k*-space of only about 2% of the Brillouin zone, and without registry occurring of any larger piece. To those inured in metal physics the overwhelming temptation, at this point, has been to turn to some density wave reconstruction of the original Fermi surface, driven either by charge, or spin, or stripe superlattices, or whatever. After all one has the ARPES "Fermi arcs" to conjure with. Accordingly we have witnessed a whole series of band-folding scenarios pass by. These invariably have however in one way or another lost sight of the true experimental circumstance we meet with in the pseudogap regime of underdoped HTSC systems. Let us take as an example the 1D-stripe scenario developed by Millis and Norman [49]. Because the Doiron-Leyraud work [18c] on oscillations in $R_H$ pointed out that the Hall coefficients are negative in these UD materials under the given conditions, some negatively signed bit of folded F.S. of appropriate area must necessarily dominate the source of the experimental signals. But where does this ascribed negative bit of F.S. derive from? It comes from the high mass, saddle region of the parent band structure, precisely where the chronic scattering was destroying quasiparticle coherence. In conflict with this prescription note that while showing the anticipated size and sign, this said bit of 'F.S.', as assayed via the quantum oscillation analysis, would manifest an effective mass which at $2m_e$ is only half that found for highly OD Tl-2201 [17], where the correlation barely had begun to be injected. Each of the proposed scenarios contains some equally suspect feature; however I do not at this point wish to pursue them all individually. I prefer to break away to present an avenue of interpretation that is quite different, one dependent upon the most mobile electrons, not the least. Moreover it is not a *k*-space-based argument, but a real space argument, one centred upon the diagonal 'rivers of charge' of the 2D-striped environment, that I have earlier set out as characterizing the underdoped systems [10,11]. This will provide an altogether more local and robust way of matching the observations than could ever be obtained in the current circumstances from a *k*-sensitive density-wave recasting of the Fermi surface geometry.

The fields employed in the new experiments (on underdoped material) of around 50 tesla are greater than $H_{irr}$ though still here below '$H_{c2}$', especially as extended into the Nernst region with underdoped material [50]. Thus we are in the region in which the fluxons are not strongly pinned but will distribute much more uniformly. With magnetic fields ~10 gauss we were able directly to observe in Bitter decoration experiments the fluxon arrays there take up regular spacing $S$ of around 1.5 μm [51]. Now with fields $5 \times 10^4$ times larger we should, with $B \propto 1/S^2$, be looking at fluxon spacings some $2 \times 10^2$ times smaller, *i.e.* $S \sim 7$ nm. At $p = 1/8$ the 2D stripe domain has edge $D = (8/\sqrt{2}) \times 4$ Å, *i.e.* 22 Å or 2.2 nm [10]. Accordingly, with the above fluxon spacing (≈ 3.2 domain lengths), around 1 in 10 domains will contain a fluxon core. Note such a fluxon commensuration number m ($\equiv S^2/D^2$) holds direct resonance with the observed 'order' of the oscillation peaks stationed in the dHvA traces near 60 T. The 'fundamental' field (m=1) under application of *reciprocal* extrapolation then comes to stand at around 600 T [18]. This is quite a small value for dHvA work, it being equivalent as pointed out above to ~ 2% of the BZ.

Before looking further at the numbers we first must consider the basics of what seems to be unfolding here. It would appear that, analogous to the set Fermi surface cross-sectional area within *k*-space of the conventional dHvA ascription, we now have in real space a fixed, *p*-determined, array of stripe domain boundaries, and past these, as the magnetic field is ramped up and down, progressively transfer the fluxons. Recall the fluxon core size is controlled by the coherence length, and for HTSC materials it may be as small as 15 Å, *i.e.* below the relevant domain sizes here. It is not unreasonable then to assume that fluxons will prefer to sit centrally within a domain, *i.e.* integrally, so that one witnesses successive magnetic free energy turning points as the fluxon number density is made to transit through, for example, the numerical commensurations 12 fluxons per 12 domain span, 11 per 11 domain span, 10 per 10, and so on, as H is ramped up.  In this process the area allotted to each fluxon grows per 'click' by a single diagonal stripe domain. One can anticipate many derivative properties such as the magnetization and the transverse Hall coefficient then to oscillate in step with the adjusting domain number counts per fluxon. Besides the magnetic aspect remember we remain below $H_{c2}$ and that at low temperatures induced supercurrents will flow around the metallic stripe boundaries where the local pairs are stable. Hence one may expect the bulk resistivity will oscillate too, even if one resides only in the Nernst region. Real space dictated oscillations not so dissimilar to the above have been recorded from the metallic edge states of quantum dot arrays constructed from materials that in the bulk state are semiconductors [52]. Considering the preceding scheme to be of potential merit, we below shall undertake a more detailed examination of the actual experimental data to see if we are able to confirm this story-line.

First it is appropriate to point out some severe difficulties are in fact posed by the current data towards any standard quantum oscillation interpretation, even were one to presume the metal examined to be quite conventional. Notably with neither Y-124 nor YBCO$_{6.5}$ do the measured fundamental "dHvA/SdH" frequencies and associated *k*-space areas express compatibility with Luttinger's theorem should the above measured pocket prove the sole *type* of pocket to exist. If just one such pocket per zone were present, it would be too small, whilst with four such pockets, it would be too big to match the number of *p*-type carriers that stoichiometry and the customary hole-type treatment of HTSC materials dictate; namely $p=0.11$ in the Y-124 case and $p=0.10$ for YBCO$_{6.5}$. If by contrast there were to be electron pockets in addition to hole pockets, as could happen with a Fermi surface reconstruction event, then further problems arise in relation to the specific heat [8]. The electronic specific heat implicit in such a case would automatically become larger than the observed value [18d], unless some carriers were to go unregistered through localization. Actually the Hall coefficient observed at low temperatures and high fields (viz. 30 mm$^3$ C$^{-1}$) provides a close match in magnitude for just a *single* pocket of the indicated size [18d]. The most striking fact though, as pointed out earlier, is that the Hall sign has under the experimental conditions crossed over into becoming *negative*. It was this latter observation which when one resorts to a more complex Fermi surface reconstruction scenario leads to problems over the sourcing of a piece of F.S. of such high mobility that it can dominate both the Hall and the SdH/dHvA responses. There is one strong indication, in fact, that this negative sign to $R_H$ at low $T$ in underdoped material issues not from Fermi surface governed density-wave formation but from stripe formation within the current, inhomogeneous, mixed-valent chemistry. This is provided by the observation that even in low fields for LBCO($p=1/8$) a comparable swing in sign of $R_H$ is closely coupled there with the LTO-to-LTT structural phase transition [18d,10]. Now diffraction measurements, whether by neutrons or electrons, have established stripe order becomes so strong in this latter material as to freeze quasi-statically, restraining $T_c$ there to uniquely low values. It therefore would appear that the negative response Hall developed in Y-124 and YBCO$_{6.5}$ comes as the strong applied magnetic field strengthens their striping tendency. From the inelastic neutron work of Lake *et al* on LSCO [53] it is known that a magnetic field drives up the l.t. magnetic spin gap aspect to the domain interiors, thereby increasing differentiation of the latter from the domain boundaries, the rivers of charge. The stripe carriers are no longer controlled in their response then to a magnetic field by the Fermi geometry, they simply are electrons and behave classically. This ought to come as a bit of a relief to Fermiologists as the Fermi surfaces of Y-123 and Y-124 are significantly different in their form and number of bands, and so why should the oscillation data they supply be so alike? As the data are independent of basal field direction they have without doubt to be associated with the planes, but one final feature affirms Fermiolgy is not the answer here; the new data support no *c*-axis dispersion whatsoever, in spite of the smallness of the claimed pockets. They exhibit the sec$\theta$ tilt behaviour of strict two-dimensionality.

Let us see now then just how closely the real space model is able to provide a match to the UD oscillation data. We shall start with a straightforward case based on $p=1/8$, the individual domains being here of edge $D = 8/\sqrt{2}.a_o$, the latter set diagonally within a square supercell of side $8a_o$ [10,11]. For the fundamental oscillation, appropriate to applied field $B_1$, we request that the fluxon density, in units of the domain area $D^2$, be equal to unity; *i.e.* $B_1 D^2/\Phi_o = 1$: this yields $B_1 = 434$ tesla. In a reduced field $B_m$, for which m vortices in the diluted vortex lattice now occur distributed over an area $(mD)^2$ (with a stripe crossing-point defining the real space origin), the associated field $B_m$ will be down on $B_1$ by the factor $1/m$; *i.e.* at these geometric coincidences $B_m \propto 1/m$. These 'high-order' fields $B_m$ and their reciprocals $1/B_m$ are evaluated in the leading section of table 1 for the first ten periods in sequence m. They relate to allowing this running number m of fluxons in the vortex superarray, of linear dimension mD, integrally to ratchet up as the field strength declines down through the listed $B_m$. While these calculated 'data points' are in the general region occupied by the actual oscillatory data, the gradient of the plot given in figure 2 is clearly too steep. What is more there appear to be something like twice as many experimental oscillations as are being calculated here. Now the above calculation was not made of course for the real doping level either for Y-124 or YBCO$_{6.5}$. Consider first the 123 material over which there is less argument as to what the correct $p$ value might be. With a $T_c$ of 57 K the YBCO$_{6.5}$ sample clearly is appreciably below the $T_c$ plateau imposed upon the Y-123 $T_c(p)$ plot by the special doping value of $p=1/8$. There exists in fact general consensus that a value of $p=0.100$ is appropriate for YBCO$_{6.5}$ (see 'methods' in [18c]), and the above fields $B_m$ accordingly will need to be rescaled by $(8/10)^2$ in order to relate to the associated $10a_o$ supercell (see table1, section 3). This automatically is going to lead to a $1/B_m$ vs. m plot which actually is even steeper now than that for $p = 1/8$ (one is required to look to reduced m to accommodate the new $1/B_m$ set). Quickly however it is noticed that the new intervals in $1/B_m$ do now follow the experimental data precisely *if* we consider half-intervals too (defined by half-integer values of m). Figure 3 shows the situation for $p = 1/10$, where we make comparison to the YBCO$_{6.5}$ dHvA data secured by Jaudet *et al* [18e]. The longest continuous trace that they present is their fig.3, reproduced now in the Appendix. We observe there that in fact it is points regularly phase shifted from the *minima* in the magnetic torque trace which perfectly match the $10a_o$ $1/2$-integer prescription. The oscillations from left to right in fig.3 of [18e] extend from (10) to (17), these running numbers being quoted here in brackets to designate that they now refer to half-integer sequencing (*i.e.* to 2m). The same association holds for the more limited set of dHvA magnetization results shown by Sebastian *et al* [18f] in their fig.4.

Two questions arise: why do half-integers seem to feature here, and why does the sequencing require counting back not always on the peaks of the various experimental signals but often elsewhere in the cycle, to reveal the 'fundamental' field, equal above to 553 tesla? Before these questions are answered it is best to examine how the situation unfolds for Y-124.

As stated there is some dispute as to what the appropriate value of $p$ actually is for YBa$_2$Cu$_4$O$_8$. I much disagree with the value of 0.14 adopted by LeBoeuf *et al* [18d] and Bangura *et al* [18b], and favour a considerably lower value, in fact somewhat lower than 0.125. This opinion is based on two counts. In YBa$_2$Cu$_3$O$_7$ the maximal $T_c$ of 92 K arises in material for which $p \approx 0.16$. There, with its stoichiometry-dictated *average* Cu valence of $2^1/_3$, one third of a hole in total has to pass to the chain Cu to leave behind just $1/_6$ of a 'hole' (re $d^9$) per planar Cu site. In YBa$_2$Cu$_4$O$_8$, with an average Cu valence now of only $2^1/_4$, an analogous transfer of one quarter of a hole in total passing to the two chain Cu's would leave behind just $1/_8$ of a hole per planar Cu site. But the hole transfer is likely in fact to be a little less since the electronegativity difference between the planes and chains is not so substantial. Thus a planar 'hole' count of $p=1/_9$ seems not unreasonable. Recall that an appreciable underdoping for Y-124 is indeed signalled by its very large $T_c$ pressure coefficient [54], $T_c$ being readily pushed up to 108 K. Such pressure coefficients arise only with significantly underdoped material (as for YBCO$_{6.5}$). Y-124 is to be considered then as positioned somewhat below the $T_c$ plateauing associated with $p=1/8$ in Y-123, *etc.*. Accordingly in table 1 section 2 one will find the comparable values of $B_m$ and $1/B_m$ relating to a $9a_o$ stripe superlattice. Possibly the action here of the strong applied magnetic field is to impose such a commensurate structure upon Y-124: LeBoeuf *et al* indeed report a sharp discontinuity in $R_H$ at 40 T [18d, fig.3c]. Once more it is evident the evaluated reciprocal fields $1/B_m$ in section 2 of the table match the 124 data (e.g. Yelland *et al* [18b,fig 2]) rather closely, though again only after being called upon to make the half-integer accommodation, as is apparent from figure 4. The fundamental field $B_{(1)}$ this time emerges from the domain model as 686 tesla, as against 556 tesla for $p = 1/10$ YBCO$_{6.5}$. The experimentally adduced values, over optimistically, were quoted as 660±15 tesla [18b2, Yelland *et al*] and 540±4 tesla [18e] respectively. Our calculated $1/B_m$ repeat *separations* match the data even better (see appendix). The field and temperature dependencies of the oscillatory signal *amplitudes* are ascribable to the level of long range order attainable in the vortex array. Under the experimental conditions one is just emerging from the pinned vortex regime around 40T.

So how might we now view the above half-integer behaviour with regard to the basal areas responsible for generating the oscillatory field traces. One straightforward interpretation would be that it represents a projection effect from an ABAB two-sheet stacking sequence of the charged domain boundaries, their displacement directed at 45° to the *b*-axis orientation of the chains. The threading of the fluxons through the sample as a whole then will be expressed by cross-sectional areas just half of what they would be were the domains and their boundaries directly to superpose between successive sheets. Note that, simple though it may be, I do not in fact advocate this solution. It relates to current circuits divided between sheets for which there is not going to be ready charge transfer, with serious impact upon mean free path maintenance of the appropriate level. Moreover it breaks symmetry as we move to area-defining oblongs, and the 45° orientation of the stacks inevitably would lead to a chaotic jumble of twins. Now from low temperature, small angle neutron diffraction scattering (SANS) studies on the best bulk samples it is known that the orientational characteristics of the vortex array are fairly stable. Even more significantly it is well established that at higher fields the diffraction spotting reveals the fluxon array to undergo transfer from being hexagonal to becoming square [55]. I immediately took this fact to signal the fluxon locations becoming constrained to the striping geometry. The authors of the SANS work presumed the striping was 1D, and thus they were not drawn to such a line of interpretation. At the time of the finding in 2002, my own view of 2D striping was that the orientation of the square domains was axial [56], and hence I did not perceive at that point that, due to the LSCO LTO $\sqrt{2}a$ structuring, the reported orientation of the square array of SANS spots in [55] was actually 'diagonal': precisely as would come to match my revised appreciation of the stripe domain geometry made in 2005 [10]. Such accommodation of the fluxon array to the striping is not the only feature of relevance here. In addition to the square order emerging only at raised fields (>0.8 tesla), the material in which it arises is underdoped, or at least in the case of Gilardi's paper for LSCO *p*=0.17 [55] having a *p* value below $p_c$. These are just the conditions under which low temperature striping becomes best organized, and accordingly most constraining upon the geometry of the traversing fluxon array. There is in fig. 2 of [55] no indication, observe, of any twinning, even with the LTO structure.

Given these circumstances, if we are not to favour a layer stacking sequence origin to account for the appearance of the half-integers in the present oscillation work, what might be the alternative? The effect in the stacking proposal was to gain a fixed set m of fluxon-encompassing, stripe-bordered regions which was twice as numerous as had been anticipated. But the same effect may of course be secured, not by halving the incremental areas, but by doubling the relevant flux quantum. If the imposed flux density is divided into quanta of $h/e$ rather than $h/2e$, all the fields $B_m$ will then need doubling to attain any desired fluxon running number, so shedding the perplexing half integers, just as manifest on figures 3 and 4. This suggestion has the advantage of being very 'clean', and of leaving the rivers of charge directly superposed between successive sheets. It is a solution that one should regard as not too unheralded, given the fact that we are dealing with a local pair superconductor and with an array of strongly anisotropic vortices tightly set about by the stripe boundaries. We already have the precedent of abnormal fluxons in the beautiful manufactured boundary experiments of Tsuei *et al* [57] that served to confirm the effectively *d*-wave nodal character of the HTSC superconductors. In the presently considered geometry we see the diagonal 2D striping array is set at an angle of 45° to the axial direction and accordingly runs in the nodal direction of the superconductivity.

How now might one cope with the second query above regarding the observation (evident in the data sets reproduced in the Appendix) that the model does not uniformly relate to one common feature between the various different oscillatory traces; say the peaks, troughs, etc.. This variety of observed phase shifts necessarily must reflect not only the different forms of experimental routine actually pursued, but also the physical character of the stripe domain interior – and in particular with dHvA how it relates to magnetism. In my analysis of the latter condition, upon appeal to the magnetic circular dichroism results of Kaminski *et al* [58], I emerged with the in-plane circulatory spin patterning within the domains displayed in figure 2b of [11]. Such an outcome has acquired support since from Fine [59] through his direct analysis of the neutron diffraction data, from Di Matteo and Norman [60] via analysis of the X-ray circular dichroism results, and from Shi *et al* [61] following further infra-red Hall, Faraday rotation and circular dichroism experimentation. It is envisaged that well within the domains the spins are settling into spin-gapped, RVB-type singlet coupling [53], but that close to the domain periphery the spins become canted, conforming to the bounding stripes under the magnetostrictive effects of the lattice developing there with the valence segregation and the Jahn-Teller effect. These canted spins then are to be viewed as responsible for the weakly magnetic conditions reported on now by Fauqué *et al* [62] and by Y. Li *et al* [63] in underdoped YBCO and HgBCO respectively: the moments upon being averaged per Cu site amount there to just ~ 0.1 μB.

We are finally in a position now to consider how we are to 'reference' the matter of phasing. In the customary Lifshitz-Kosevich formulation of the oscillatory magnetization the form used is $\sin(2\pi.F/B + \phi)$. In these terms the traces appearing in the Appendix for $\Delta\rho/\rho$ from Bangura *et al* [18b1], for the inductance signal from Yelland *et al* [18b2], for the oscillatory torque data from Jaudet *et al* [18e], and for $-\Delta R_H$ from Doiron-Leyraud *et al* [18c], are respectively of phasing $\phi$ = 0, $\pi/2$, $\pi$ and $-\pi/2$. A second detail of phasing is how to position the stripe superlattice itself relative to the real space origin: does one select a stripe or a domain centre to be so placed? The simplest choice would be the former because we have used the stripes themselves for counting purposes. But does that choice then mesh in meaningfully with the experimental observations? It would seem so. $\Delta f$ in Yelland *et al*'s experiment is $\propto dM/dB$, making $M$ itself maximal at $\phi = \pi/2$, where the flux vortex will sit central to the domain and the magnetic $^9Cu_{II}^0$ sites then will be maximally affected. The Jaudet *et al*'s torque phasing of $\pi$ results from $\tau = -(\Delta \mathbf{M}) \times \mathbf{B}$ in the presence of antiferromagnetic coupling. Maxima in $|\Delta\rho|$ (or rather minima in longitudinal mobility) are to be expected under phasing for which each flux line coincides with a transverse stripe, supplying $\phi = 0$ for Bangura *et al*'s trace. The $\phi = -\pi/2$ phasing present in Doiron-Leyraud *et al*'s trace is because they follow $-\Delta R_{Hall} \propto -\Delta p$, the effective change in content of responding holes in underdoped material. It would seem then that the choice made above is the appropriate one, and that these various differences in trace phasing also look capable of being satisfactorily embraced within the proposed stripe domain/vortex array modelling.

The actual cyclical profile of the experimental oscillatory signal is finally, one might note, neither decaying monotonically, as Yelland *et al* point out in [18b], nor is it of simple harmonic form. There seems to be a tendency to a more complex oscillation profile, perhaps of saw-tooth character. The analysis made by Sebastian *et al* [18g] would support this upon reading their third Fourier peak to be the third harmonic. One finds it to be considerably stronger than the second harmonic. Unfortunately Sebastian *et al* have interpreted their third peak as evidence of an independent orbit, the outcome of being unable to specify their fundamental closely enough due to the short field span that their experiment covers.

In closing this section I would like to add a more formal statement of the law of dilution of the vortex array as $H$ falls which has been employed above:

    Take a square superarray of vortices, side m$D$, and consider m vortices in this area.
    Let the system dilute to (m+1) vortices in square superarray of side (m+1) linear units, D,
        or (m+1)$^2$ areal units, $D^2$.
    The incremental area is 2m+1  or  m + (m+1) .
     Each of original m vortices has increased its areal footprint by 1 (stripe domain) to (m+1),
        and we also have embraced the extra vortex with this same footprint (m+1).
    In this way have gained recurrent sequencing of (m+1) vortices in an (m+1)$D$ superarray.

To distribute these vortices uniformly and as widely as possible it would appear the solution is to place just one vortex in each column and one in each row. Figure 5a shows the solution for m=13. It is seen that the vortex array itself is square also in this case, but actually canted relative to the diagonal directions of the stripes, unlike the vortex superarray of side 13$D$. Figure 5b shows an equivalent vortex footprint for m=13, which being lower in symmetry is not to be favoured over the form given in figure 5a. A further paper will present the likely development in footprint form which occurs with vortex compaction as $H$ changes. The SANS results of [55b] guide these deliberations below 7 tesla.

It is hoped that the detailed exposition given above will wean people away from viewing the experimental results from these underdoped HTSC systems as standard quantum oscillations betokening the existence of a Fermi surface and Fermi liquid. That presumption surely was misguided from the start, and it is trusted the present work has opened up an altogether more appropriate and potentially exciting scenario to cover the observations.

§5. Summary.

With §1 a brief survey was provided of the negative-$U$ setting of the boson-fermion crossover scenario for HTSC systems, and of the marked differences existing between the nodal and antinodal conditions. These latter have great implications for the resonance between the local pairs and induced BCS-like Cooper pairs, including how the screening conditions set up at the various doping levels in any given HTSC mixed-valent system afford direct control over the maxima reached in $T_c(p)$.

§2 traces in detail the development of the abnormal electrical and optical properties of HTSC systems as 'hole doping' is decreased below $p$ = 0.3. For the latter $p$ value the position of the Landau Fermi liquid behaviour is fairly standard if highly correlated. Below $p$ = 0.3 there arises a marked growth in scattering and a steady shift towards incoherence. Apportionment of the unusually strong scattering into elastic and inelastic, isotropic and anisotropic contributions is interpreted in terms of $e/e$-to-$b$ and $e$-on-$b$ events after close examination of the $T$- and $p$-dependent evolution of the AMRO, Seebeck and Hall data sets.

§3 takes a more detailed look at new high quality $\rho(T,p)$ resistivity data and analysis from LSCO, and at the sharp changes coming at $p$ = $p_c$ (=0.18). The nature of what arises there as one meets with strong incoherence and the pseudo-gap condition is described in terms of the crossover model.

§4 follows up on the consequences of such an interpretation and reformulates the setting of the high-field oscillatory magnetization, resistivity and Hall data obtained recently from UD material in terms *not* of the standard, quantum oscillation, $k$-space response of a fairly normal Fermi liquid, but of a real-space-based action that is much more appropriate to the UD condition. Success here is demonstrated to revolve around the 2D striping supported in such mixed-valent materials, and seemingly enhanced in definition in the strong applied magnetic field. The model is one of response to the free energy changes arising as the unpinned vortex array slips discretely across the stripe array. A very close match to the experimental data can be secured *provided* that the flux quantum in question in this ring-threading process is not $h/2e$ but $h/e$. The observations are related back to Forgan and colleagues' observations on UD material of symmetry conversion from hexagonal to square in the vortex array geometry as the magnetic field is increased. Those skeptical of the proffered outcome and requiring theoretical justification of the $h/e$ quantum of flux in question should seek out the paper by Vakaryuk to

which my attention has just been drawn in the current issue of PRL [64]. This would look to relate to just the type of situation described above, although it is not clear to me how far the fact we are not dealing with point bosons, particularly in the Nernst region above $T_c$, might be significant here.

**Postscript;** I have just received communication from Prof. A. S. Alexandrov that in fact he has already released a Fast Track paper in April relating these oscillatory observations not to quantized Fermiology, but, as with my own paper, to a real space origin [*J. Phys.: Condens. Matter* **20** 192202 (2008)]. Somehow I had not noticed this work, which, like my own, arises from a disquiet with the Fermi surface controlled interpretation of events and with a conviction that real-space pairing is behind HTSC superconductivity. The present scenario is, though, significantly different from that of Prof. Alexandrov in that I believe the pairing mechanism to be more electronically based in the negative-*U* resonant crossover procedure, rather than it being the outcome of bipolaron condensation. More specific to the present oscillation problem, I look to the sourcing of the real space periodicity with which the vortex array interacts to come not from the checker-boarding but from the 2D striping (see [10] for distinction).

   Finally I would draw the reader's attention to the new release from Audouard, Jaudet and coworkers [66] claiming to supply 3D extension to their previous dHvA deliberations [18e]. Note that as far as my own interpretation is concerned their two new subsidiary frequencies simply are those relating to a certain amount of $11a_o$ and $9a_o$ stripe superlatticing about the mean $10a_o$ state discussed above: N.B. for the $11a_o$ case $B_{(1)} = 2\times(8^2/11^2)\times 434$ T = 459 T – they extract a $B_{(1)}$ of 453 T. It is remarkable that the stoichiometry in 123-YBCO$_y$ can be homogenized even as well as this. In 124 of course we have a fixed stoichiometry, which above we treated as giving $p = \frac{1}{9}$. The likelihood is that this equivalence is not exact, similarly leading to subsidiary oscillations. Note in the new 123 work [66] that the fraction of a ~140×140×40 μm sample to support the $11a_o$ trace ($p$=0.091) is observed to grow as the empirical overall oxygen stoichiometry drops from $y$ = 6.54 to 6.51. It is essential now to examine the $y$ = 6.45 samples known to exist, where $p \sim \frac{1}{13}$.

**Acknowledgements;** I would like to thank Nigel Hussey and colleagues for continued discussions on HTSC matters and in particular for advancing to me a copy of their latest work, upon which §3 is centred, prior to its submission for publication.

**Table 1.** (p.14)

Model values (see text) for oscillatory repeat features to be expected with "striping" superlattices of $8a_o$, $9a_0$ and $10a_o$ due to interaction with vortex array. In the modelling of [10,11] the 'striping' is in fact two-dimensional with the stripes or 'rivers of charge' running in the diagonal directions.

- For $p = 1/8$:   Supercell edge = $8a_o$

    Average basal lattice const. $a_o$ = 0.3855 nm.

    ∴ diagonal domain edge $8a_o/\sqrt{2}$ = 2.1810 nm, (= $D$)

    and diagonal domain area   = 4.7568 nm$^2$. (= $D^2$)

    For 'fundamental' oscillation appropriate to applied field $B_1$ require fluxon number

    density, in units of domain area $D^2$, to equal *unity*, i.e. $B_1 D^2 / \Phi_o = 1$,

    taking as $\Phi_o$, the flux quantum, $h/2e$ = 2.067 x 10$^{-15}$ Wb or 2067 tesla nm$^2$.

    Hence here   $B_1$ (= $\Phi_o/D^2$) = 2067/4.7568 = 434.5 tesla, – and   $B_m = B_1/m$.

- For the case appropriate to $p = 1/9$, where dealing with domain of edge $9a_o/\sqrt{2}$, scale by $(8/9)^2$

    to get   $B_1$ = 343.3 tesla.   (or $B_{(1)}$ = 686.2 tesla, see text).

- For case appropriate to $p = 1/10$, where dealing with domain of edge $10a_o/\sqrt{2}$, scale by $(8/10)^2$

    to get   $B_1$ = 278.1 tesla.   (or $B_{(1)}$ = 556.2 tesla, see text).

- Y-123$_y$, $a_{av}$, see  R K Siddique 1994 *Physica C* **228** 365 (fig.6) ; Y-124, $a_o$ = 0.3841 nm, $b_o$ = 0.3871 nm.

| Dom. edge Calc$^d$.order m | $8a_o/\sqrt{2}$ $B_m$ tesla | $8a_o/\sqrt{2}$ $1/B_m$ | $9a_o/\sqrt{2}$ $B_m$ tesla | $9a_o/\sqrt{2}$ $1/B_m$ | $10a_o/\sqrt{2}$ $B_m$ tesla | $10a_o/\sqrt{2}$ $1/B_m$ | Data order | $m^{exp}$ |
|---|---|---|---|---|---|---|---|---|
|   | *869.0* | *0.00115* | *686.2* | *0.00146* | *556.2* | *0.00180* | (1) | ½ |
| 1 | 434.5 | 0.00230 | 343.1 | 0.00291 | 278.1 | 0.00359 | (2) | 1 |
|   | - |   | *228.7* | *0.00437* | *185.4* | *0.00539* | (3) | 1½ |
| 2 | 217.3 | 0.00460 | 171.5 | 0.00583 | 139.0 | 0.00719 | (4) | 2 |
|   | - |   | *137.2* | *0.00729* | *111.2* | *0.00899* | (5) | 2½ |
| 3 | 144.8 | 0.00690 | 114.4 | 0.00874 | 92.7 | 0.01078 | (6) | 3 |
|   | - |   | *98.0* | *0.01020* | *79.5* | *0.01256* | (7) | 3½ |
| 4 | 108.6 | 0.00921 | 85.7 | 0.01166 | 69.5 | 0.01438 | (8) | 4 |
|   | - |   | *76.2* | *0.01311* | *61.8* | *0.01618* | (9) | 4½ |
| 5 | 86.9 | 0.01151 | 68.6 | 0.01457 | 55.6 | 0.01798 | (10) | 5 |
|   | - |   | *62.4* | *0.01603* | *50.6* | *0.01978* | (11) | 5½ |
| 6 | 72.4 | 0.01381 | 57.2 | 0.01749 | 46.4 | 0.02157 | (12) | 6 |
|   | - |   | *52.8* | *0.01894* | *42.8* | *0.02337* | (13) | 6½ |
| 7 | 62.1 | 0.01611 | 49.0 | 0.02040 | 39.7 | 0.02517 | (14) | 7 |
|   | - |   | *45.7* | *0.02186* | *37.1* | *0.02697* | (15) | 7½ |
| 8 | 54.3 | 0.01841 | 42.9 | 0.02332 | 34.8 | 0.02877 | (16) | 8 |
|   | - |   | *40.4* | *0.02477* | *32.7* | *0.03056* | (17) | 8½ |
| 9 | 48.3 | 0.02071 | 38.1 | 0.02623 | 30.9 | 0.03234 | (18) | 9 |
|   | - |   | *36.1* | *0.02769* | *29.3* | *0.03416* | (19) | 9½ |
| 10 | 43.5 | 0.02301 | 34.3 | 0.02915 | 27.8 | 0.03596 | (20) | 10 |
| Re | - |   | YBa$_2$Cu$_4$O$_8$ $p$ = 0.111 |   | YBa$_2$Cu$_3$O$_{6.5}$ $p$ = 0.100 |   |   |   |

**Captions**

**Figure 1.** (*p.7*)

Schematic rendering of the way in which the strong anisotropy existing in the mean free path $l(\mathbf{k}_F)$ controls the magnitude and sign of the Hall coefficient in overdoped HTSC materials. The latter sign is dictated by the Ong integral $-\int_{FS} d\mathbf{l} \times \mathbf{l}$ taken around the Fermi surface [30]. The signs of the incremental contributions are here set by the sense locally of the circulation of the cross-product; counter-clockwise circulation leads to negative contributions to $R_H$, clockwise to positive. The construction shows how the net sign of the integral depends on the difference in area of the two types of closed segment generated within the full circuit traced out by $l(\mathbf{k}_F)$. The figure is produced for the situation appropriate to OD-LSCO, which has the form of FS indicated and where a small mean free path in the axial directions converts to a much larger mean free path in the diagonal 45° directions. The outcome is a positive Hall coefficient in spite of the Fermi surface here being closed about the centre of the BZ, not the corner. The special locations for $l(\mathbf{k}_F)$ marked P to T divide the $l(\mathbf{k}_F)$ picture up into areas designated $\alpha$ to $\theta$ (and match points p to t for $\mathbf{k}_F$ around the Fermi surface). q is the point of curvature inflection on the FS and tangential point Q on $l(\mathbf{k}_F)$ marks the associated maximum to angle $\phi$ for points within the first half quadrant. Signs are inserted both by $\mathbf{dl}$ and by the vector cross product.

Where $\mathbf{dl} \times \mathbf{l}$ is positive from P to Q, the integral itself is given by $+(\alpha+\beta+\gamma+\delta)$,

" is negative from Q to S, " by $-(\gamma+\delta+\varepsilon)-(\beta+\eta+\zeta)$,

" is positive from S to T, " by $+(\beta+\eta+\gamma+\theta)$.

So for entire repeating quadrant the net " is given by $+(\alpha+\beta+\gamma+\theta)-(\varepsilon+\zeta)$,

- or taken over all four quadrants, the difference in areas of the counter-clockwise minus clockwise enclosed areal segments — and here negative, making $R_H$ positive.

**Figure 2.** (*p.15*)

The set of calculated values of $1/B_m$ versus integers m ($= S^2/D^2$) for case of $p = {}^1/_8$ and square superlattice $8a_o$. Following the 'stripe' modelling of [10] the stripes themselves are diagonal in orientation and define here domains of edge $8a_o/\sqrt{2}$. Note as $S^2 = mD^2$, $m(S^2) = (mD)^2$; i.e. m fluxons will thread a square set of $m^2$ stripe domains. The fundamental field $B_1 = 434$ T.

**Figure 3.** (*p.15*)

Reworking of figure 2 for $p = {}^1/_{10}$ and striping superlattice period of $10a_o$, appropriate to case of YBCO$_{6.5}$. The crosses and circles relate to the data cycles recorded by Doiron-Leyraud *et al* [18c;$R_H$] and Jaudet *et al* [18e;$\tau$]. There are twice as many oscillations, (m), as were first expected, m, (see text), but the match of the calculated $B_{(m)}$ values to the full integer and half-integer data set is excellent. The fundamental value of $1/B_{(1)}$ inverts to a field of 556 T.

**Figure 4.** (*p.16*)

Similar plot to that of figure 3 but now for $p$ value of $^1/_9$ and a striping superlattice period of $9a_o$, appropriate to case of YBa$_2$Cu$_4$O$_8$. The crosses and circles relate to the data cycles recorded by Bangura *et al* [18b1;$\Delta\rho/\rho$] and Yelland *et al* [18b2;inductance]. Here fundamental value for $1/B_{(1)}$ of 0.00146 inverts to a field $B_{(1)} = 686$ T.

**Figure 5.** (*p.*18)

Two vortex tiling solutions for the case m = 13. The small squares (side *D*) are the stripe domains (two-dimensional and in the diagonal/nodal orientation). 13 vortices lie within the square super-array 13*D* x 13*D*. The vortices in this particular case themselves form a square array of side √13.*D* in (3,2) rotation to the nodal direction. Case (a) is of higher tile (i.e. vortex footprint) symmetry than case (b), and so more likely to be adopted. In moving from commensuration m=12 to 13 (to 14), every vortex footprint augments by 1 stripe domain during the dilution step. For all comparable tilings the most compact geometric forms always are to be favoured.

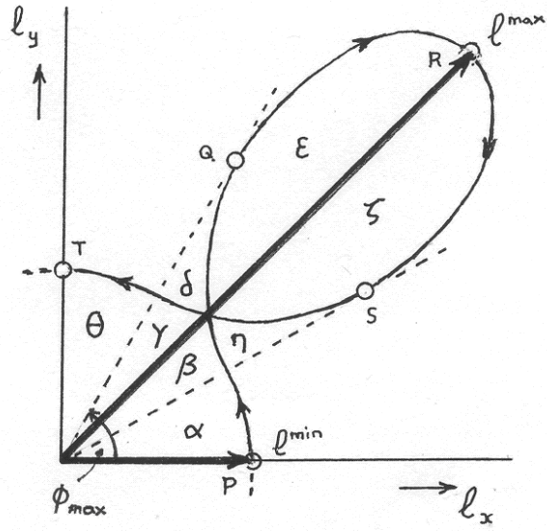

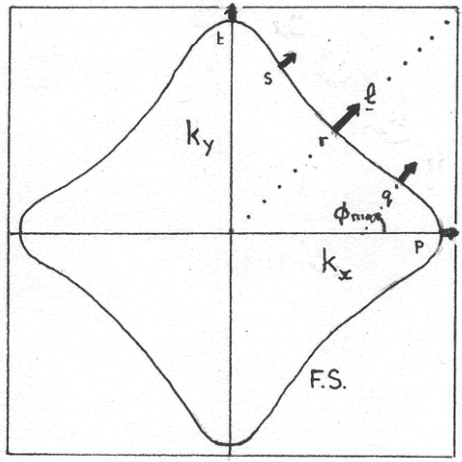

f1

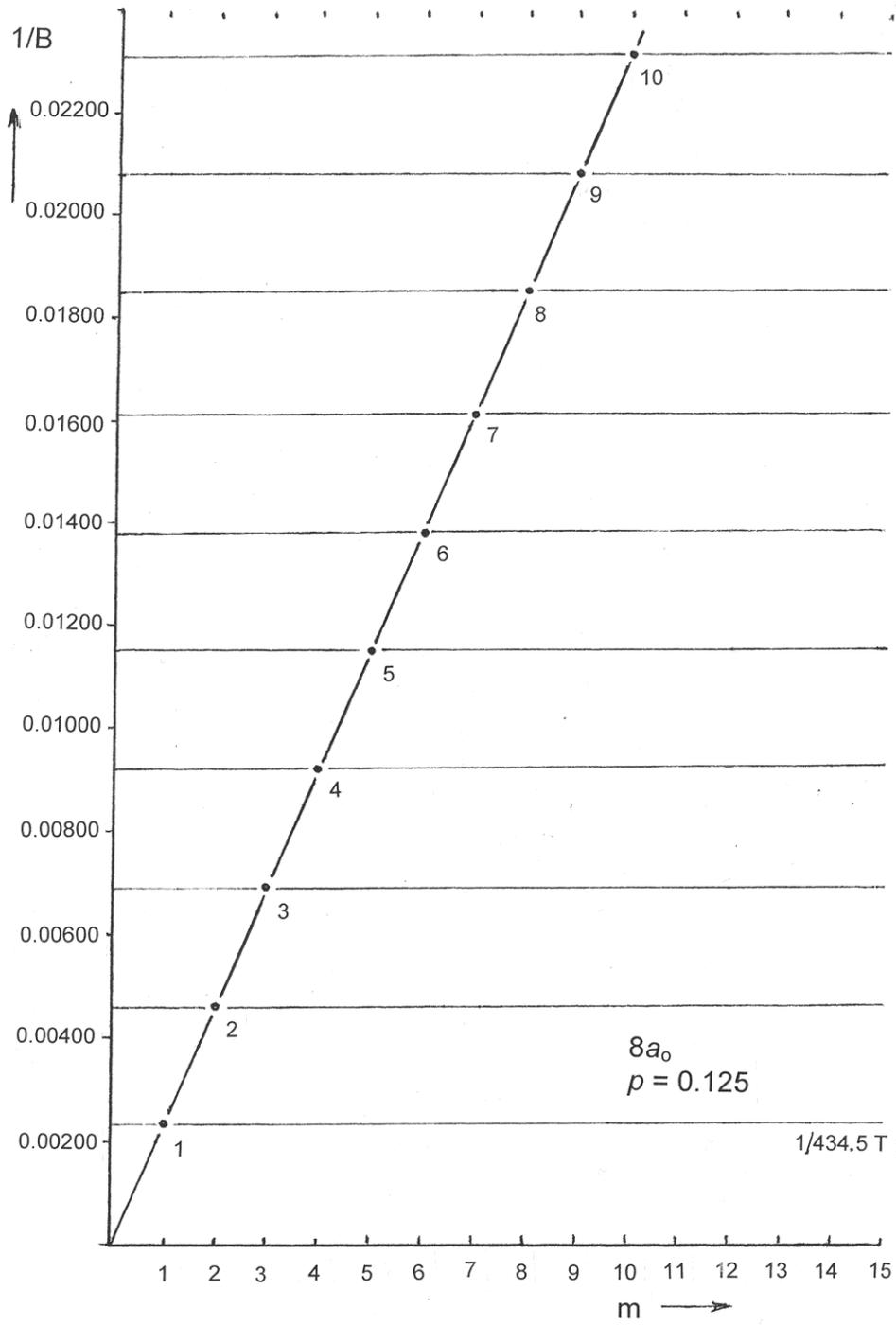

f2

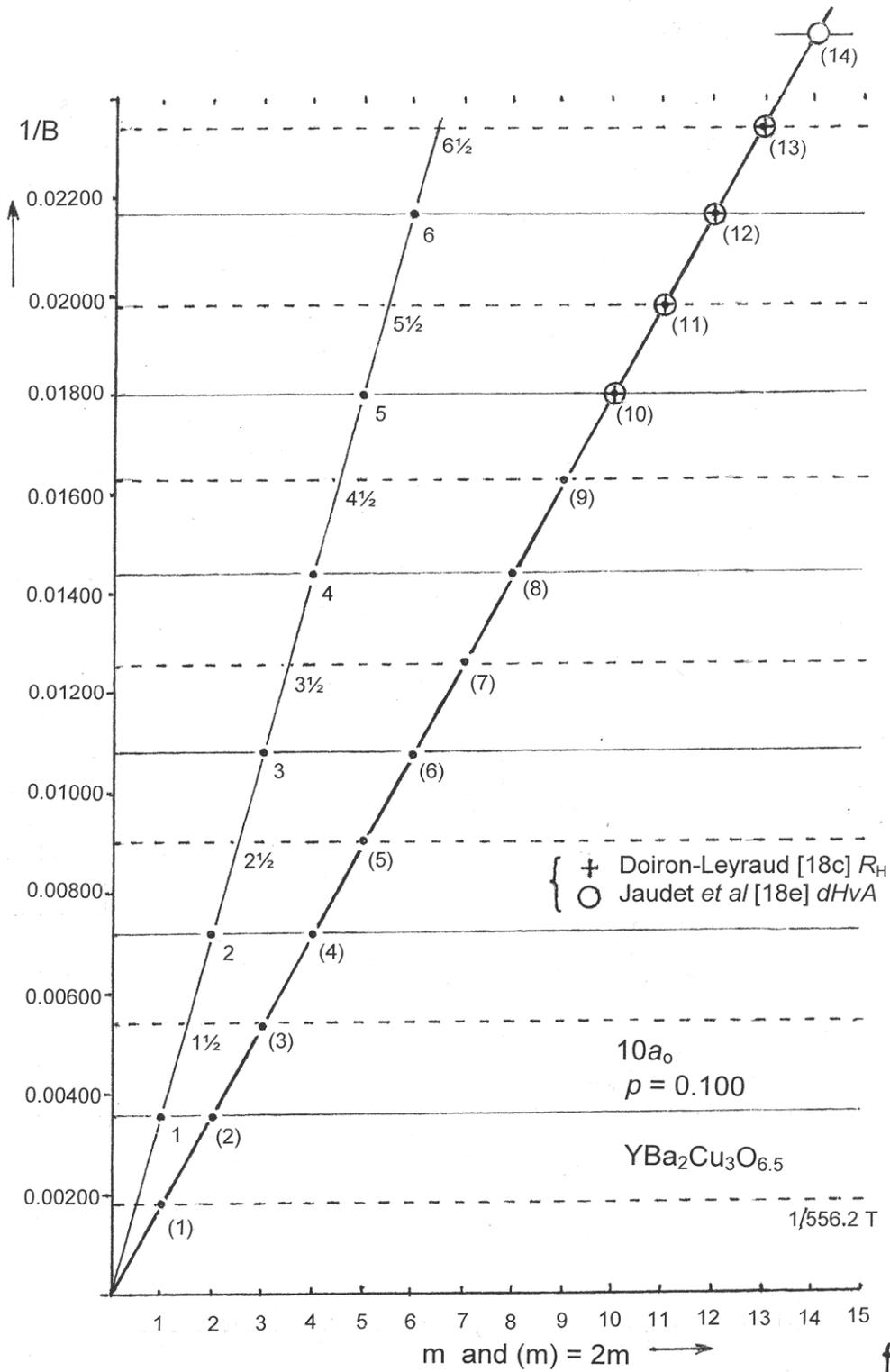

f 3

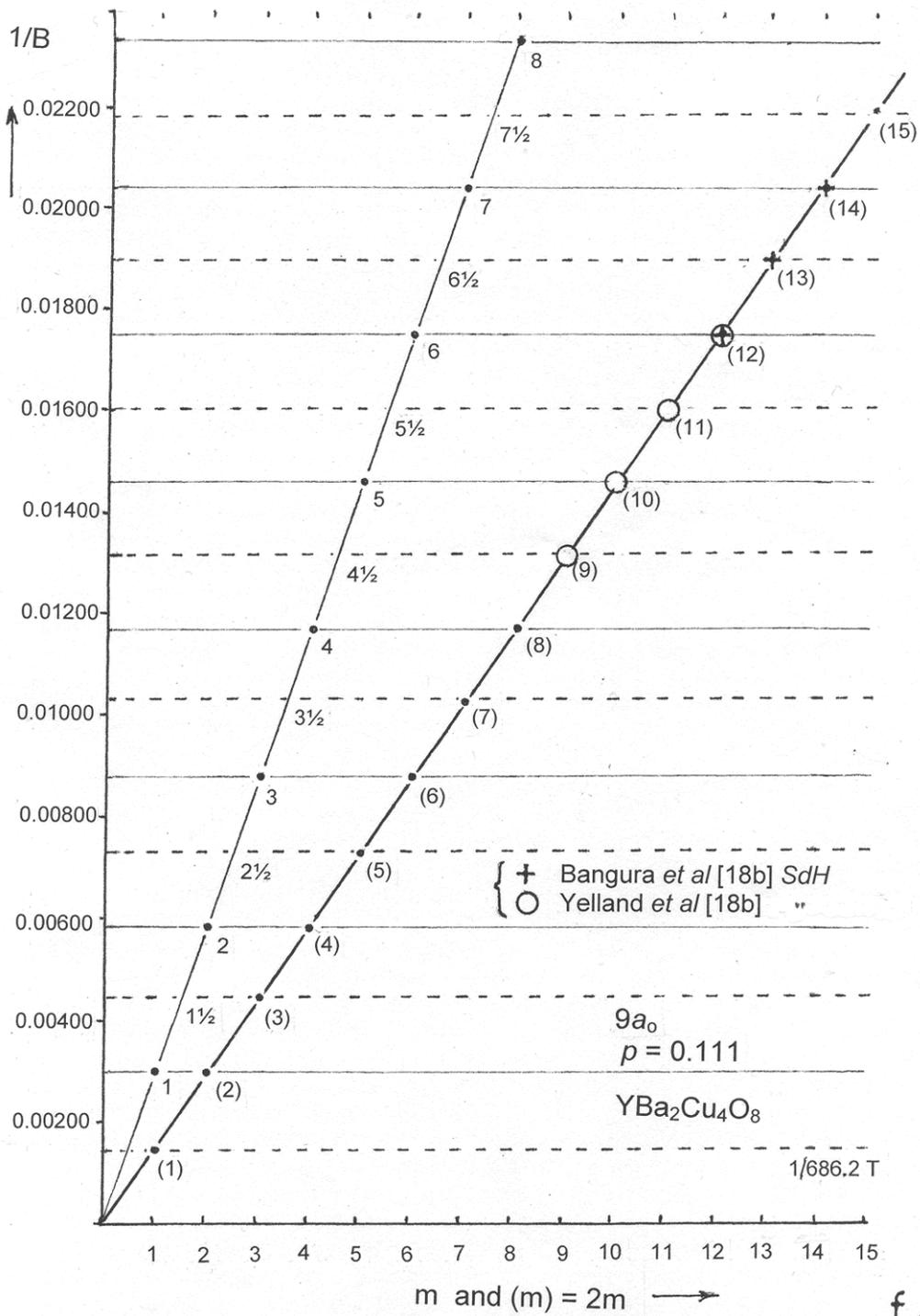

f4

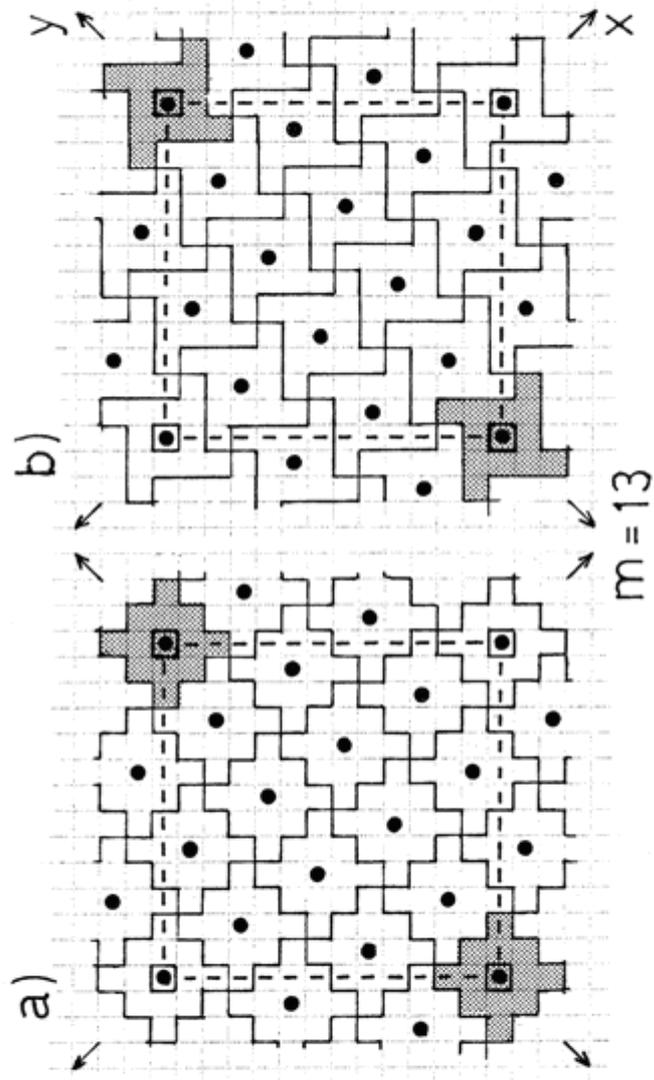

m = 13

**Appendix**

Indication is made directly on the published data of the level of match for the two underdoped HTSC materials to the oscillation periods secured when employing the present stripe-domain/vortex-array modelling as opposed to the customary Fermi liquid dictated quantum oscillation interpretation. Figure A1 is for $YBa_2Cu_3O_{6.51}$, and the data come from Doiron-Leyraud *et al* [18c;fig.3a] and Jaudet *et al* [18e;fig.3a]. The present evaluation takes the hole doping there to be 0.100 supporting a $10a_o$ domain superlattice. Figure A2 is for $YBa_2Cu_4O_8$, and the data come from Bangura *et al* [18b1;fig.2] and Yelland *et al* [18b2;fig.2]. The present evaluation takes the hole doping here to be 0.111, supporting a $9a_o$ domain superlattice.

The various phase shifts evident for these data traces are addressed in the text.

The matches presented require one to adopt as flux quantum in the present circumstances a value of $h/e$ rather than $h/2e$, as is highlighted in the text and as very recently broached in the theoretical literature by Vakaryuk [64].

YBa$_2$Cu$_3$O$_{6.5}$   $p = 0.100$   $10a_o$

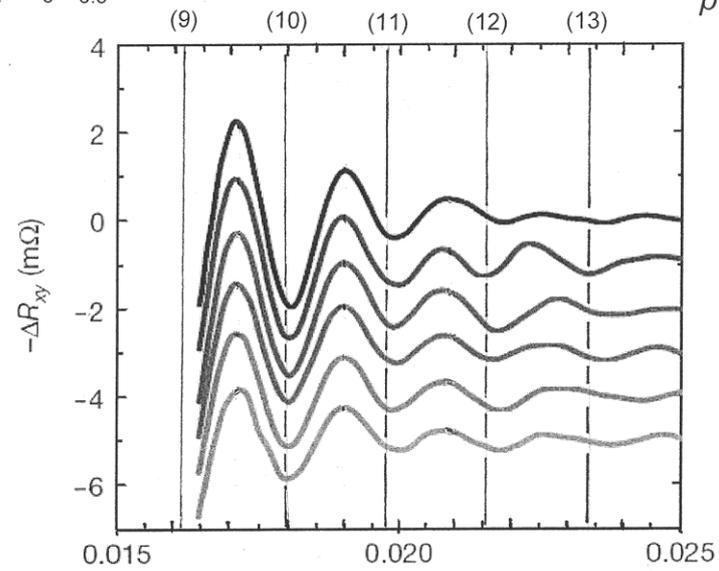

• Doiron-Leyraud [18c] $R_H$

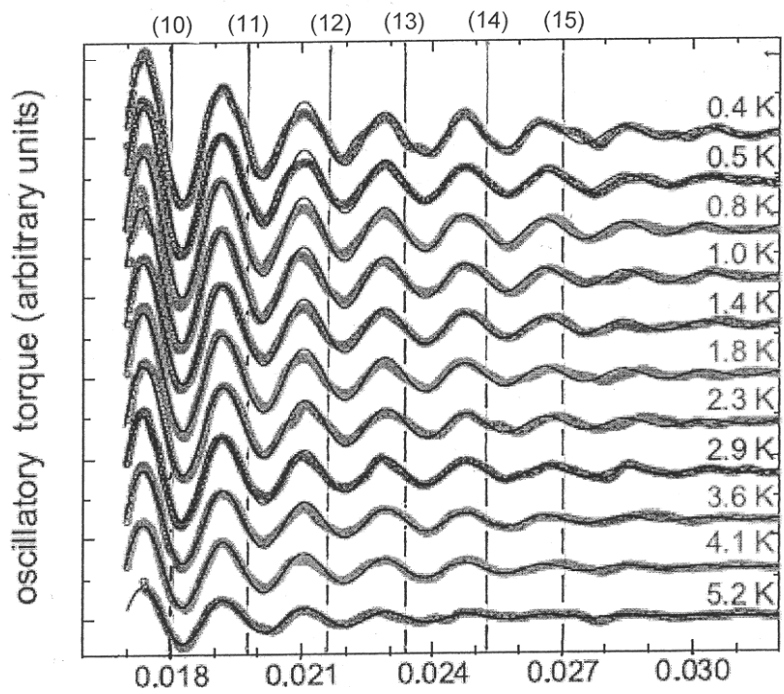

• Jaudet et al [18e] dHvA

fA1

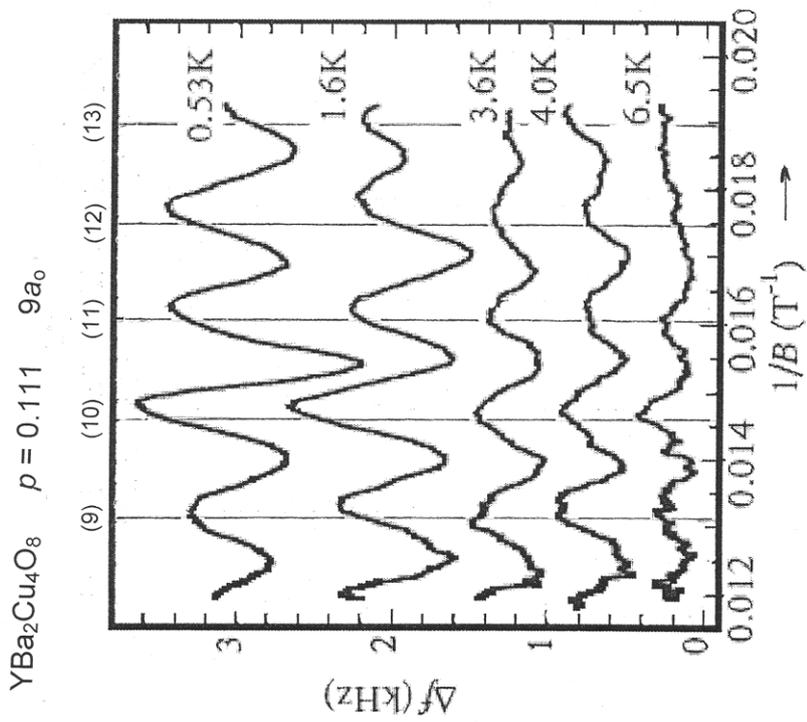

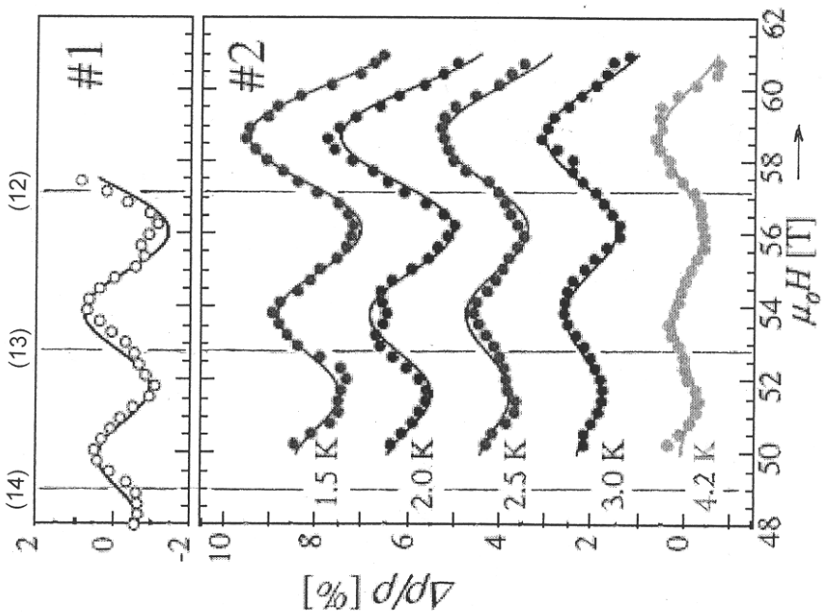